\documentclass[conference]{IEEEtran}
\IEEEoverridecommandlockouts
\usepackage{cite}
\usepackage{amsmath,amssymb,amsfonts}
\usepackage{algorithmic}
\usepackage{graphicx}
\usepackage{textcomp}
\usepackage{xcolor}
\usepackage{physics}
\def\BibTeX{{\rm B\kern-.05em{\sc i\kern-.025em b}\kern-.08em
    T\kern-.1667em\lower.7ex\hbox{E}\kern-.125emX}}
\begin{document}

\title{Enhancing the Performance of DeepReach on High-Dimensional Systems through Optimizing Activation Functions \\
{\footnotesize }

}

\author{\IEEEauthorblockN{Qian Wang\textsuperscript{*}}
\thanks{* These authors contributed equally to this work.}
\IEEEauthorblockA{\textit{University of Southern California}\\
Los Angeles, USA \\
pwang649@usc.edu}
\and
\IEEEauthorblockN{Tianhao Wu\textsuperscript{*}}
\IEEEauthorblockA{\textit{University of Southern California} \\
Los Angeles, USA \\
wutianha@usc.edu}
}

\maketitle

\begin{abstract}
With the continuous advancement in autonomous systems, it becomes crucial to provide robust safety guarantees for safety-critical systems. Hamilton-Jacobi Reachability Analysis is a formal verification method that guarantees performance and safety for dynamical systems and is widely applicable to various tasks and challenges. Traditionally, reachability problems are solved by using grid-based methods, whose computational and memory cost scales exponentially with the dimensionality of the system. To overcome this challenge, DeepReach, a deep learning-based approach that approximately solves high-dimensional reachability problems, is proposed and has shown lots of promise. In this paper, we aim to improve the performance of DeepReach on high-dimensional systems by exploring different choices of activation functions. We first run experiments on a 3D system as a proof of concept. Then we demonstrate the effectiveness of our approach on a 9D multi-vehicle collision problem. 

\end{abstract}


\section{Introduction and Motivation}
With the continuous advancement in autonomous systems, it's becoming more and more significant to ensure safe operations of these systems. Providing robust safety guarantees for autonomous systems still faces lots of challenges due to various reasons such as inherent uncertainties in the systems and their operating environments. Recently, one popular approach that aims to address this problem is Hamilton-Jacobi (HJ) Reachability analysis, which has been widely applied in robotics to compute the \textit{Backward Reachable Tube (BRT)} of a dynamical system. To illustrate the concept of BRT, consider a simplified model where two vehicles are heading toward each other on a narrow path. In this example, BRT is the set of all states of the two vehicles that a collision becomes inevitable, despite of the best possible efforts of both drivers to avoid the collision. A more precise mathematical formulation of this problem will be given in a later section. 

HJ Reachability analysis offers a mathematical framework to compute the BRT of a dynamical system by converting the reachability problem into an optimal control problem. It can handle general nonlinear dynamics and non-deterministic uncertainty in the system under state constraints and control bounds. Many related toolboxes have been developed to solve HJ reachability: DeepReach (Python), HelperOC (Matlab), BEACLS (C++), etc. However, the computation of BRT becomes highly intractable when dealing with high-dimensional systems when using grid-based methods because of the \textit{curse of dimensionality}. This drawback limits their applicability to most practical robotic systems that can't be modeled in low dimensions.

DeepReach is a deep learning-based method that approximates the value function through a deep neural network (DNN) \cite{b1}. One huge advantage of DeepReach is the computation required for obtaining the value function doesn't scale exponentially with the dimension of the system as for grid-based methods. Instead, it only scales with the underlying complexity of the value function, which makes it a promising tool to approach high-dimensional reachability problems. Currently, DeepReach deploys a sine activation function in the DNN based on recent research results that show periodic, sinusoidal activation functions can well represent signal's derivatives \cite{b2}. Therefore, sine better models the gradient of the value function and leads to a more accurate approximation. While this approach works very well on low-dimensional systems, it remains a challenge to accurately approximate the value function for high-dimensional systems. The main tradeoff is between the difficulty of the optimization problem to solve and the quality of the gradient during the training of the DNN. Sine activation function captures better gradients, but at the same time, makes the optimization process much harder. On the contrary, ReLU results in a much easier optimization process, but a worse representation of the gradients because its first derivative is piece-wise constant.

In this paper, we propose using a combination of Sine and ReLU activation functions in the DNN to better approximate the value function for higher dimensional systems. We first run sanity checks on low-dimensional systems to make sure that by incorporating ReLU activation, we can still obtain a generally sound BRT with respect to the analytical solution \cite{b3}, though a larger MSE is expected when compared with the state of the art model that only uses Sine activation. Then we experiment with a 9-dimensional multi-vehicle collision system. To quantitatively measure the performance of BRTs, we compute their violation rate based on results from recent research \cite{b9}.

\section{Related Work}
\subsection{Reachability Analysis for Dynamical Systems}

Traditionally, BRTs for dynamical systems are obtained through the level-set approach. It discretizes the state space into a grid, then  numerically solves the partial differential equation (PDE) of the value function using existing level set toolboxes\cite{b4}. However, one main disadvantage of this approach is that the computational cost scales exponentially with the system's dimensionality. As a result, this method can be applied up to 5D systems \cite{b1}, \cite{b5}.

A recent advance in reachability analysis is DeepReach. It aims to address the aforementioned challenge by using a DNN to approximate the value function instead of directly solving the PDE. DeepReach is a promising approach toward high-dimensional reachability problems, but some challenges remain. For example, to the best of our knowledge, no existing method can accurately verify BRTs for high-dimensional systems. Nevertheless, compared with previous methods that take the union of pairwise BRTs as an approximation for the overall BRT, DeepReach is able to solve the original high-dimensional problem directly \cite{b1}.

\subsection{Solving PDE with Neural Network}

In reachability analysis, we formulate it as an optimal control problem, then using dynamic programming, the problem turns into solving a Hamilton-Jacobi Isaacs (HJI) PDE. DeepReach gets its inspiration from recent progress in solving differential equations using DNN \cite{b6},\cite{b7}. After training, we obtain a model that represents the value function. Given a state \textit{x} and time \textit{t}, it outputs the value function evaluated at (\textit{x},\textit{t}), denoted by \textit{V(x,t)} \cite{b1}.

\subsection{Activation Function in DeepReach}

To overcome the challenge of inaccurate representation of derivatives with smooth, non-periodic activation functions \cite{b2}, DeepReach deploys periodic, sinusoidal activation functions that have achieved very small errors on low-dimensional systems \cite{b1}. As previous experiments have shown, using the Sine activation function in the DNN achieves an MSE an order of magnitude lower than other choices such as ReLU, Tanh, and Sigmoid. The ReLU architecture fails to learn the correct value function on a 3D system. ReLU represents a nearly linear function and therefore preserves the properties of linear models that make it easy to optimize \cite{b8}, but at the cost of less accurate derivatives. It remains an open question of how to take advantage of both the efficient optimization of ReLU and the accurate  derivative representation of Sine, which we are trying to explore in this paper.

\section{Problem Formulation}
Hamilton-Jacobi Reachability analysis is a mathematical framework to obtain the backward reachable tube of dynamical systems, which can be described by a Hamilton-Jacobi equation. The problem is formulated as follows:

Consider a dynamical system described by the ordinary differential equation

\begin{equation}
    \dot{x}(t) = f(x(t),u(t),d(t))
\end{equation}

where $x \in \mathbb{R}^n$ is the state of the system, $u \in \mathbb{R}^m$ is the control input, $d \in \mathbb{R}^m$ is the disturbance, and $f$ is a smooth vector field that defines the dynamics of the system.

Suppose that we want to find the set of all initial states $x_0 \in \mathbb{R}^n$ from which we can steer the system to a target set $T \subseteq \mathbb{R}^n$ in a finite time horizon $[0, T_f]$ while minimizing a given cost function $J(x,u,d)$. This set is called the backward reachable tube of the system, denoted by $BRT(T, J)$.

We first apply the level set method and compute its value function $V(x)$. Then, we are able to recover the BRTs by taking the sub-zero level set of the value function. Formally, the Hamilton-Jacobi-Issac variational inequality involves finding a function $V(x)$ that satisfies the following PDE:

\begin{equation}
\begin{split}
\frac{\partial V(x)}{\partial t} + \min_{u(t)} \max_{d(t)} {\frac{\partial V(x)}{\partial x} \cdot f(x(t),u(t),d(t))}&=0 \\
V(x, T)&=l(x)
\end{split}
\end{equation}

The Backward Reachable Tube (BRT) is then defined as:

\begin{equation}
B R T(T, J)=\left\{x: \forall u(\cdot) \in \mathbb{U}_{t}^{T}, \exists d(\cdot) \in \mathbb{D}_{t}^{T}, \xi_{x, t}^{u, d}(s) \in T \right\}
\end{equation}

where $BRT(T, J)$ is the set of all states that can be reached in a finite time horizon $s \in [0, T_f]$ for which the agent acts optimally and under
worst-case disturbances will eventually reach the target set
T within the time horizon $[0, T_f]$. The set $BRT(T, D, u(\cdot))$ can be computed using the value function obtained by solving the HJ-PDE.

The value function $V(x)$ can be used to compute the reachable set $BRT(T,J)$ as follows:

\begin{equation}
    BRT(T,J) = \{x \in \mathbb{R}^n \mid V(x) \leq 0\}
\end{equation}

Once the value function is computed, an optimal control policy that minimizes the cost function $J(x,u)$ subject to the constraints imposed by the reachable set $BRT(T, J)$ can be obtained using the following formula:

\begin{equation} \label{eqn:5}
    u^*(x) = \arg\min_{u \in U} \max_{d \in D} { \langle \nabla V(x), f(x,u,d) \rangle}
\end{equation}

where $u^*(x)$ is the optimal control input at state $x$.

\section{Proposed Approach}
DeepReach uses a deep neural network with 3 hidden layers, each containing 512 nodes, to approximate the value function. The input layer and the 3 hidden layers utilize a sinusoidal activation function.

The choice of activation function can significantly impact a neural network's performance and training time. While the sinusoidal activation function has demonstrated superior performance, it is also more computationally demanding than the popular Relu activation function. To balance the trade-off between speed and performance, we investigate different combinations of activation functions in each layer to achieve optimal performance while minimizing training time.

We first run experiments on a 3D system as proof of concept. Then, we turn to a 9D multi-vehicle collision system, where traditional BRT solvers struggle to find a solution and analytical computation is not feasible. Therefore, our neural network lacks a formal way to test the accuracy of the results. To address this issue, we implement a method for interpreting and validating the neural network's output.


\subsection{Intertwined Activation Functions}

The Relu activation function is one of the easiest for a neural network to optimize, and a Sine function is usually harder. Therefore, we combine the efficiency brought by the Relu with the performance produced by the Sine activation function. We first fix our neural network structure to be 1x3x1 (1 input layer, 3 hidden layers, and 1 output layer same as the structure used in DeepReach. We also fix the epoch size to 120000 starting with the air3D problem and then moving on to 150000 epochs for the multi-vehicle collision problem. 

For our experiment, we treat two parameters of the neural network as variables: the number of Sine layers and the arrangement of Sine layers. We use all Sine and all ReLU layers as our baselines.





\subsection{Result Evaluation}

Validating the accuracy of the computed set is challenging, especially for high-dimensional systems where finding the ground truth of the reachable set is impossible. In this section, we propose a similar method as the scenario optimization proposed in the work of \cite{b9}.

In this method, we sample a finite number of states uniformly across the spectrum of the entire state space. We then apply optimal control $u^*(x)$ (Equation \ref{eqn:5}) and optimal disturbance $d^*$ on each sample state and generate trajectory $\xi_{x, t}^{u, d}(s)$ until $t=T_f$. We define \textbf{violation rate} as the ratio of the number of violations to the total number of sampled states. In our work, we sample 100K states in each experiment.

There are two conditions that can be treated as a violation: 

1) $$\exists x \in BRT^C \land s \in [0, T_f] : \xi_{x, t}^{u, d}(s) \in T$$

2) $$\exists x \in BRT : \forall s \in [0, T_f],  \xi_{x, t}^{u, d}(s) \notin T$$

The first condition says that any state outside of the BRT should not traverse the failure set at any time on the trajectory. The second condition says that any state sampled inside the BRT should intersect the failure set, or else will be treated as a violation. 

The second metric we use to evaluate the model is $\delta_{\tilde{V}, \tilde{\pi}}$, defined as the maximum learned value of an empirically unsafe initial state under the induced policy $\tilde{\pi}$ :

\begin{equation}
\delta_{\tilde{V}, \tilde{\pi}}:=\max _{x \in X}\left\{\tilde{V}(x, 0): J_{\tilde{\pi}}(x, 0) \leq 0\right\}
\end{equation}

where $J_{\tilde{\pi}}(x, 0)$ is the cost function associated with the trajectory obtained by using the policy $\tilde{\pi}(x, t)$ from an initial state $x$ and initial time $t=0$ until $T_f$. Here on, we use $\delta$ as a shorthand for $\delta_{\tilde{V}, \tilde{\pi}}$ for brevity purposes.

Intuitively, the violation rate measures how often the model deviates from the true BRT, and $\delta$-level measures how far away the worst violating state is from the true BRT.

\section{Case Studies and results}
\subsection{Running Example: Air3D}

We first run our experiments on a relatively simple 3-dimensional system as a sanity check for the proposed methods. The system is formulated as a collision avoidance problem between two identity vehicles. To reduce the dimensionality from six to three, the system dynamics are represented by the relative dynamics between the two vehicles as the following:

\begin{equation}
\begin{split}
&\dot{x}_1 = -v_e + v_p\cos x_3 + \omega_e x_2 \\
&\dot{x}_2 = v_p\sin x_3 - \omega_e x_1 \\
&\dot{x}_3 = \omega_p - \omega_e
\end{split}   
\end{equation}

Here, $x_1$ and $x_2$ are the relative positions, and $x_3$ is the relative heading between the two vehicles. $v_e$ and $v_p$ are constant linear velocities with a value of 0.75 m/s. $\omega_e$ and $\omega_p$ are angular velocities bounded to the range of [$-3, 3$] in radian.

The target set is defined as the set of all states where the two vehicles are within a certain distance R to each other. In this example, R is chosen to be 0.25m.

\begin{equation}
    \mathcal{L} = \{x: \lVert(x_1, x_2) \lVert \leq R \}
\end{equation}

\begin{figure}[htbp]
\centerline{\includegraphics[width=80mm,scale=0.5]{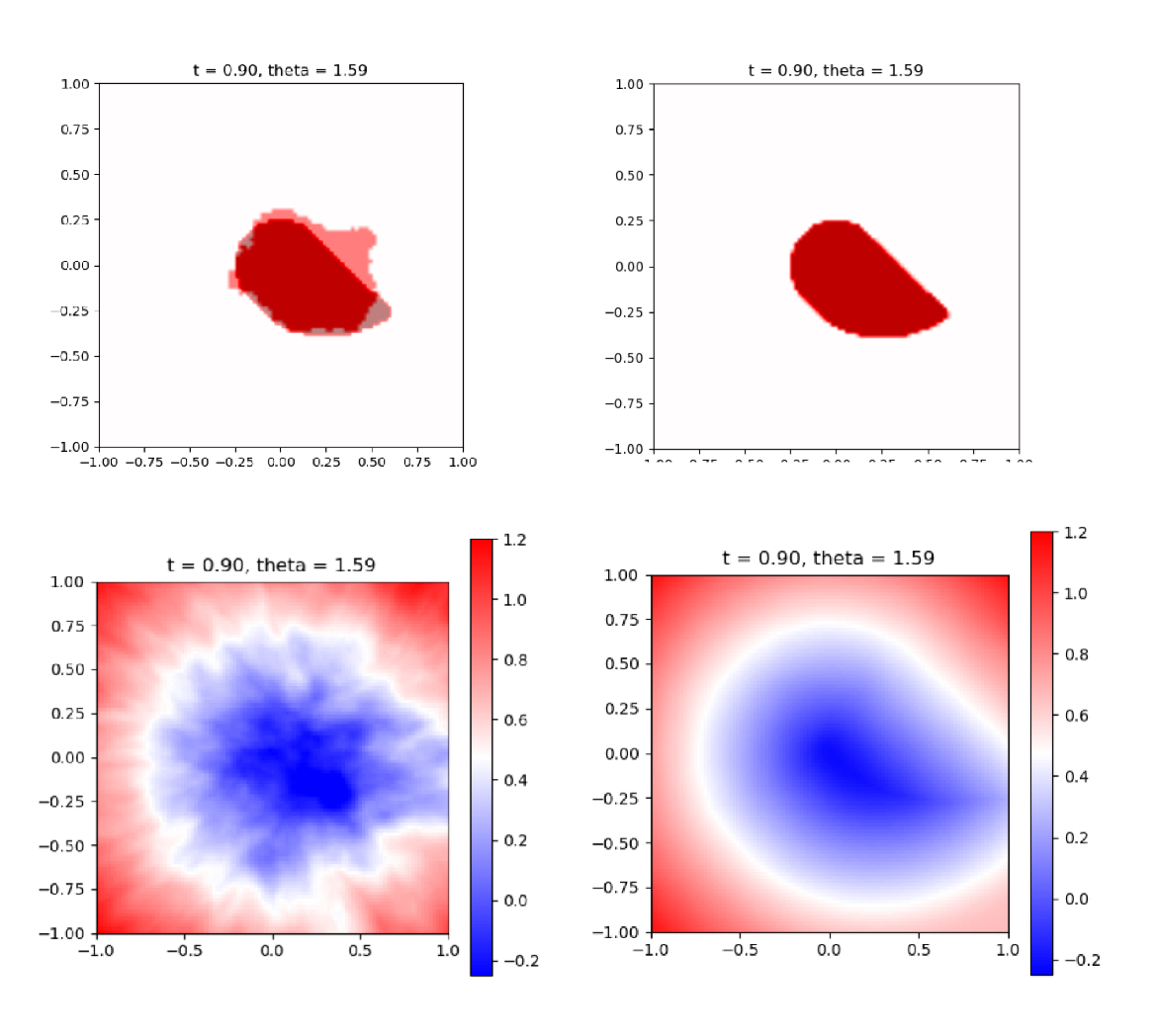}}
\caption{Baseline experiment on air3D}
\label{allrelu}
\end{figure}

Figure 1 above shows the baseline for our experiment. The left two plots are the results of using ReLU in all layers. The right two plots are from using Sine. The top two figures compare the BRTs (dark red region) we get with the ground truth BRT (grey region if not overlapped) obtained using the grid-based method. The bottom two figures are the original BRTs we learn through training, where the blue region is the sub-zero level set of the value function, which is exactly our BRT. We can observe that ReLU poorly learns the BRT while Sine is able to almost fully recover the true BRT, with an MSE on the magnitude of $10^{-4}$.

Then we run various experiments using a combination of Sine and ReLU.
The table below records some preliminary results. In the model structure row, ``r'' means we are using a Relu activation function, and ``s'' means we are using a Sine activation function. We pre-train the model for 10k iterations, followed by 110k curriculum training.

\begin{table}[htbp]
\begin{center}
\begin{tabular}{ |c||c|c|c|c|  }
 \hline
 \multicolumn{5}{|c|}{Air3D Results} \\
 \hline
 Experiment ID& exp1 &exp2&exp3&exp4\\
 \hline
 Model structure    &ssrsl&   srsrl&rrrsl&srrrl\\
 Epochs&   120000  &120000&120000&120000\\
 Training time & 1:49:03 & 1:41:57&  1:35:29&1:37:35\\
 \hline
\end{tabular}
\label{tab1}
\end{center}
\end{table}

\begin{figure}[htbp]
\centerline{\includegraphics[width=80mm,scale=0.5]{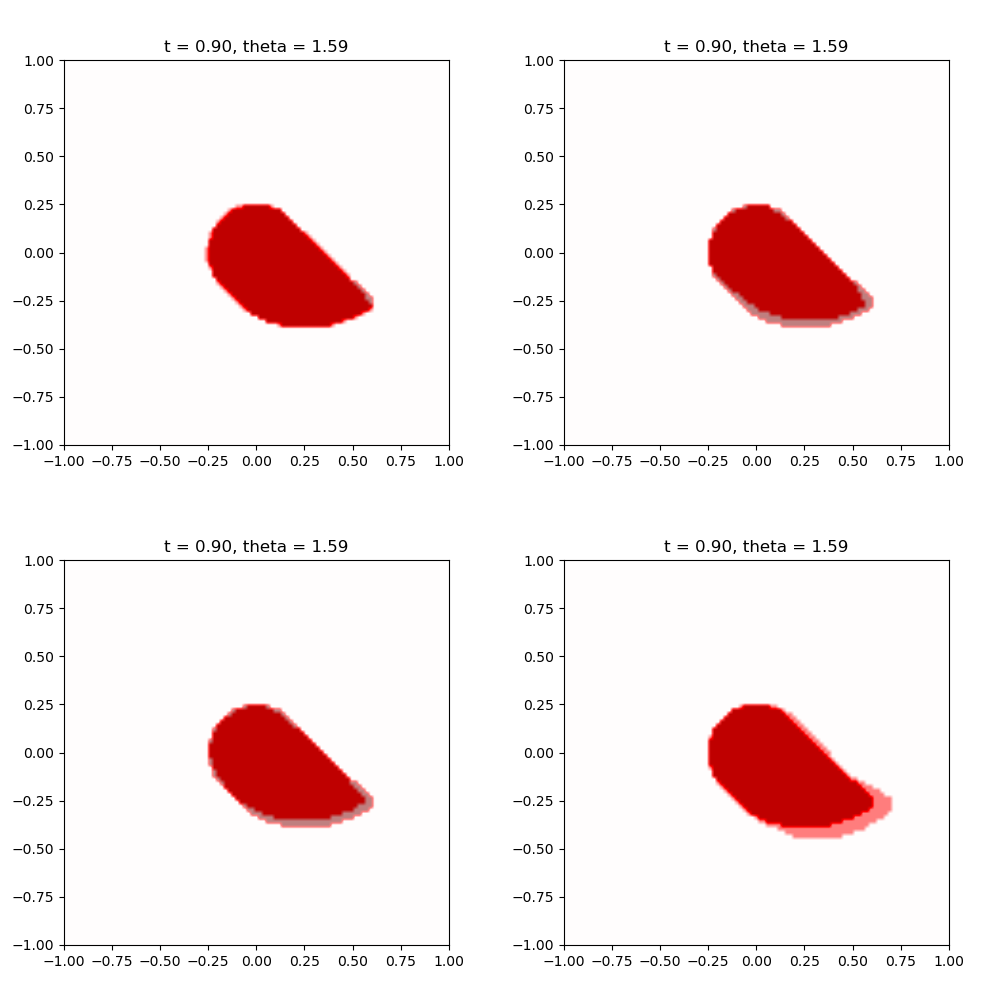}}
\caption{The slices of BRT for t=0.9 and theta=$\pi/2$ on air3D. The BRT in crimson is obtained from helperOC as the “ground truth” to compare with results from DeepReach (in light red). The four figures correspond to the four experiments we ran using different combinations of Sine and ReLU: upper left (exp1), upper right (exp2), bottom left (exp3), bottom right (exp4)}
\label{fig}
\end{figure}

There is a couple of interesting findings here. First, the training time is strongly related to the number of layers used as a Sine activation function. Second, the arrangement of the Relu and Sine has some effect on the training time as well: using Sine on the first few layers slows down training time compared to using it on the last few layers when the number of Sine layers is the same, as demonstrated in exp3 and exp4. Generally, the more Sine activation we use in the neural network, the more accurate BRTs we obtain.


\subsection{Running example: Multi-vehicle Collision Avoidance}

In the previous example, we confine the system to 3D such that it's possible to compare the results with the ground truth BRT. Now we release the dimensionality to 6D by framing the same pursuer-evader game as the joint state space between the two vehicles. Formally, the dynamics are defined as:

\begin{equation}
\begin{split}
&\dot{x}_1 = v_p \cos x_3 \hspace{3 mm} \dot{x}_2 = v_p  \sin x_3 \hspace{3 mm} \dot{x}_3 = \omega_p  \\
&\dot{x}_4 = v_e \cos x_6 \hspace{3 mm} \dot{x}_5 = v_e\sin x_6 \hspace{3 mm} \dot{x}_6 = \omega_e
\end{split}   
\end{equation}

Similar to Air3D, the control is $w_e$ and disturbance $w_p$. The target set is defined as:

\begin{equation}
    \mathcal{L} = \{x: \lVert(x_1, x_2) - (x_3, x_4) \lVert \leq R \}
\end{equation}

To work on a 9D three-vehicle collision avoidance problem, we first obtain the BRT for the two-vehicle 6D system, then take the union of the pairwise BRTs as an approximation to the solution of the overall 9D system. Notice that this approximation is often inaccurate because it doesn't take into account collisions that happened due to three-way interactions between the vehicles. We simply use this approximation as a visual evaluation metric for the performance of DeepReach on 9D systems. The 9D system is obtained by adding a new evader into the 6D problem, which adds 3 new dimensions to the dynamics.

Similar to Air3D, we run several experiments with different combinations of Sine and ReLU. Because the system has a higher dimensionality, we increase the pretraining iterations from 10k to 40k. Everything else remains the same as Air3D. The table below records the training parameters.

\begin{table}[htbp]
\begin{center}
\begin{tabular}{ |c||c|c|c|c|c|c|  }
 \hline
 \multicolumn{7}{|c|}{Multi-vehicle Collision Results} \\
 \hline
 Experiment& exp5 &exp6&exp7&exp8&exp9&exp10\\
 \hline
 Structure    &ssrsl& rsrsl&rrrsl&srrrl&ssssl &rrrrl\\
 Epochs&   150000  &150000&150000&150000&150000 &150000\\
 Time & 1:58:08 & 1:50:39& 1:44:25&1:43:31&2:05:21 &1:38:26\\
 \hline
\end{tabular}
\label{tab1}
\end{center}
\end{table}

As expected, the training time shows a linear relationship with the number of layers that use Sine. Figure 3 shows the BRTs for the base cases ssssl and rrrrl.

\begin{figure}[htbp]
\centerline{\includegraphics[width=80mm,scale=0.5]{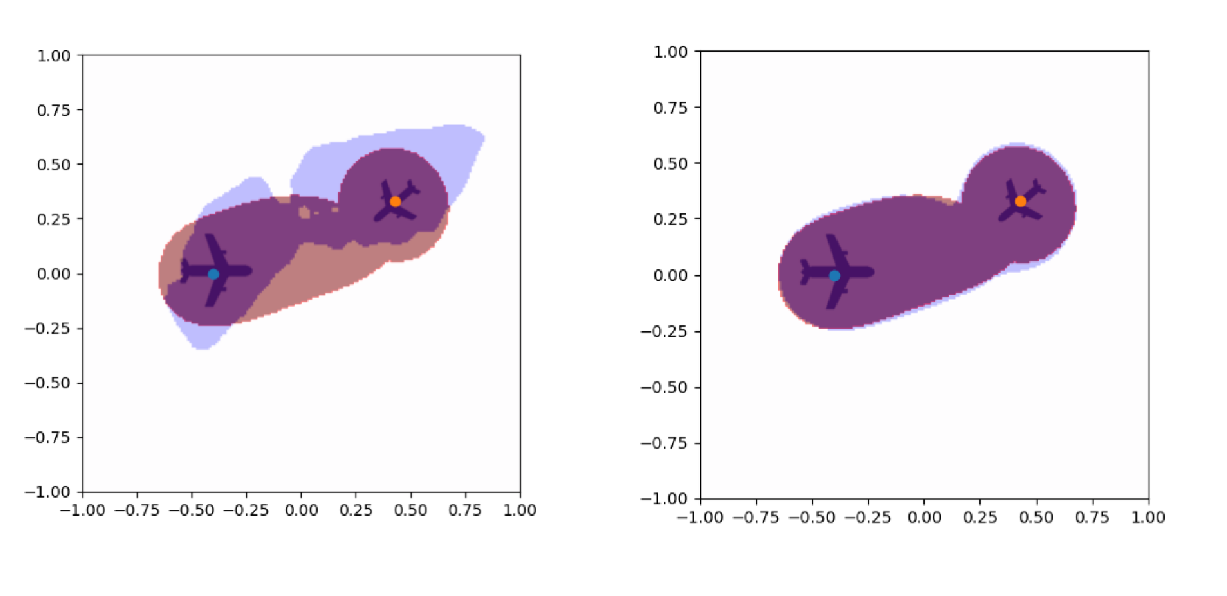}}
\caption{Base case experiments on 9D multi-vehicle collision. 
left (exp10) right (exp9). The brown region is the union of pairwise BRTs whereas the light blue region is the learned from the whole 9D system. The overlapping region between the two BRTs is purple.}
\label{allrelu}
\end{figure}

By visual inspection, we discover that using ReLU in all layers inaccurately approximated the value function as it even fails to capture a large portion of the union of pairwise BRTs. On the other hand, Sine is able to cover the whole brown region, which indicates it's probably a more accurate approximation.

\begin{figure}[htbp]
\centerline{\includegraphics[width=80mm,scale=0.5]{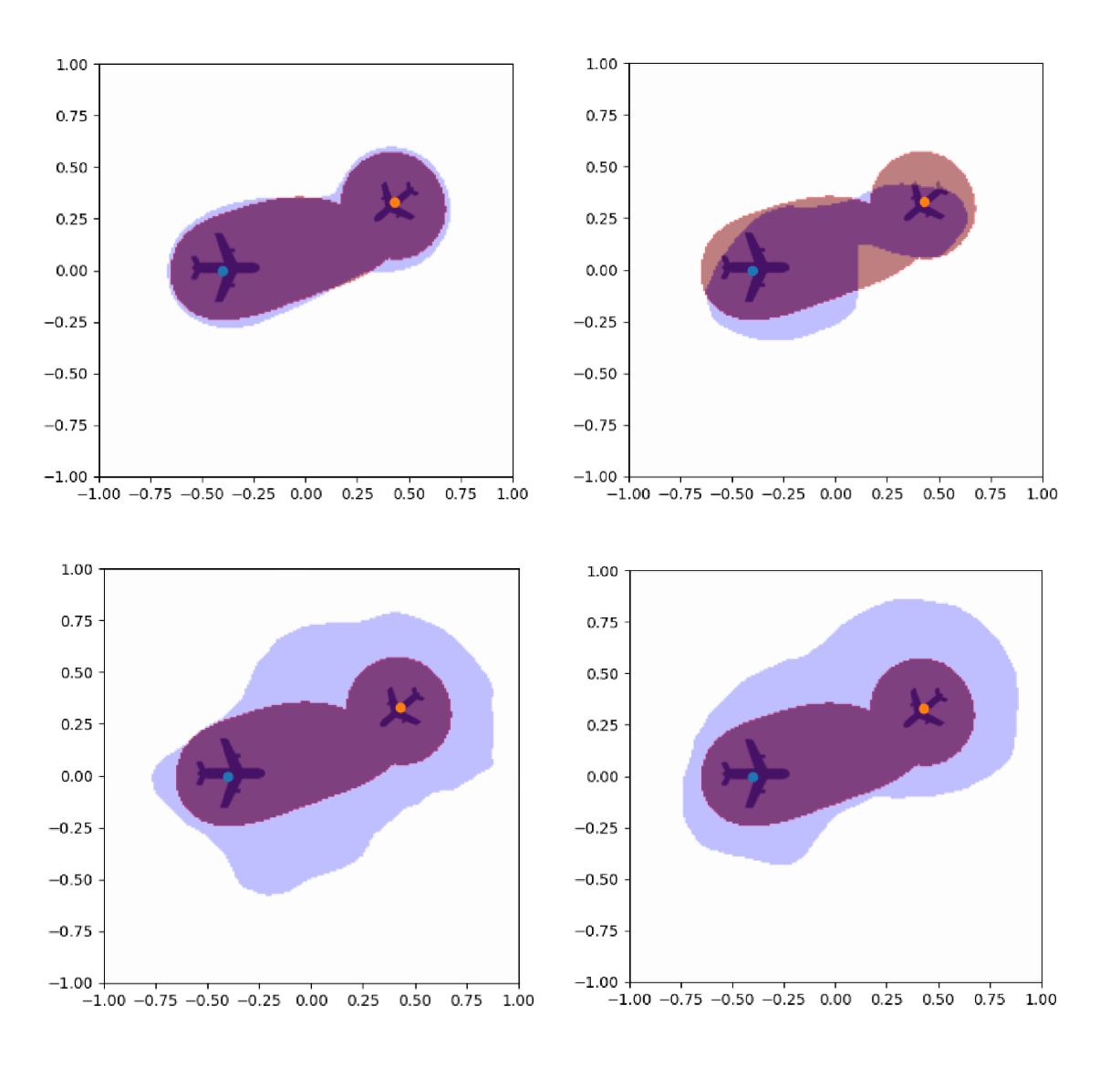}}
\caption{Upper left (exp5), upper right (exp6), bottom left (exp7), bottom right (exp8). The coloring used is the same as in Figure 3.}
\label{allrelu}
\end{figure}

Figure 4 displays the learned BRTs when using combinations of Sine and ReLU. exp5, exp7, and exp8 learn BRTs that cover the union of pairwise BRTs. BRTs from exp7 and exp8 appear to be much larger than the union of pairwise BRTs. While the figure gives us some information about the quality of the BRTs intuitively, we need to be careful about making premature conclusions about these results. The plots project the BRTs to a lower dimension for visualization purposes. So they are just slices of the entire BRTs. Even with the slices of BRTs, we are unsure about whether a larger region as in exp7 and exp8 or a smaller region as in exp5 is better. Hence, to analyze the results  more objectively, we need a method that returns numerical values about the quality of the BRTs. We defer more detailed numerical analysis to the next section.

\section{Result analysis}

With the BRTs obtained for the 9D system from the last section, we implement the verification procedure to quantify their performance.

\begin{table}[htbp]
\begin{center}
\begin{tabular}{ |c||c|c|c|c|c|c|  }
 \hline
 \multicolumn{7}{|c|}{Multi-vehicle Collision Evaluations} \\
 \hline
 Experiment& exp5 &exp6&exp7&exp8&exp9&exp10\\
 \hline
 Structure    &ssrsl& rsrsl&rrrsl&srrrl&ssssl &rrrrl\\
 Violation \% & 18.97 & 21.09 & 21.31 & 27.15 & 19.00 & 23.53 \\
 $\delta$-level & 0.441 & 0.891 & 0.803 & 0.916 & 0.550 & 0.999 \\
 \hline
\end{tabular}
\label{tab1}
\end{center}
\end{table}

We have three observations from the table:

1)
There exists a strong correlation between the number of Sine layers used and the performance of the model.

2)
Under the same number of Sine layers, the position of the Sine layer can greatly impact the performance of the model demonstrated in ex7 and exp8.

3) 
ssrsl outperforms the baseline ssssl, and srrrl performs worse than rrrrl, as shown respectively in exp 5 and exp8.

\vspace{3 mm}

exp5 potentially indicates that ReLU helps the learning of the value function. To further test our hypothesis, we run three more experiments on the 9D system with six layers. We add a ReLU layer for two experiments at different positions, and one experiment with all Sine layers. The results are shown in the following table.

\begin{table}[htbp]
\begin{center}
\begin{tabular}{ |c||c|c|c|  }
 \hline
 \multicolumn{4}{|c|}{Multi-vehicle Collision Evaluations} \\
 \hline
 Experiment& exp11 &exp12&exp13\\
 \hline
 Structure    &ssrrsl& srsrrl&sssssl\\
 Violation \% & 19.45 & 27.61 & 18.43 \\
 $\delta$-level & 0.747 & -0.085 & 0.352\\
 \hline
\end{tabular}
\label{tab1}
\end{center}
\end{table}

sssssl achieves the best result so far among all experiments, with a violation rate of 18.43. But adding one more ReLU layer doesn't generate a smaller violation rate as we expected. Surprisingly, srsrrl gives the worst violation rate.
To better understand how different activation structures affect the accuracy of BRTs on high-dimensional systems, we plot the results for experiments on the 9D system.

\begin{figure}[htbp]
\centerline{\includegraphics[width=80mm,scale=0.5]{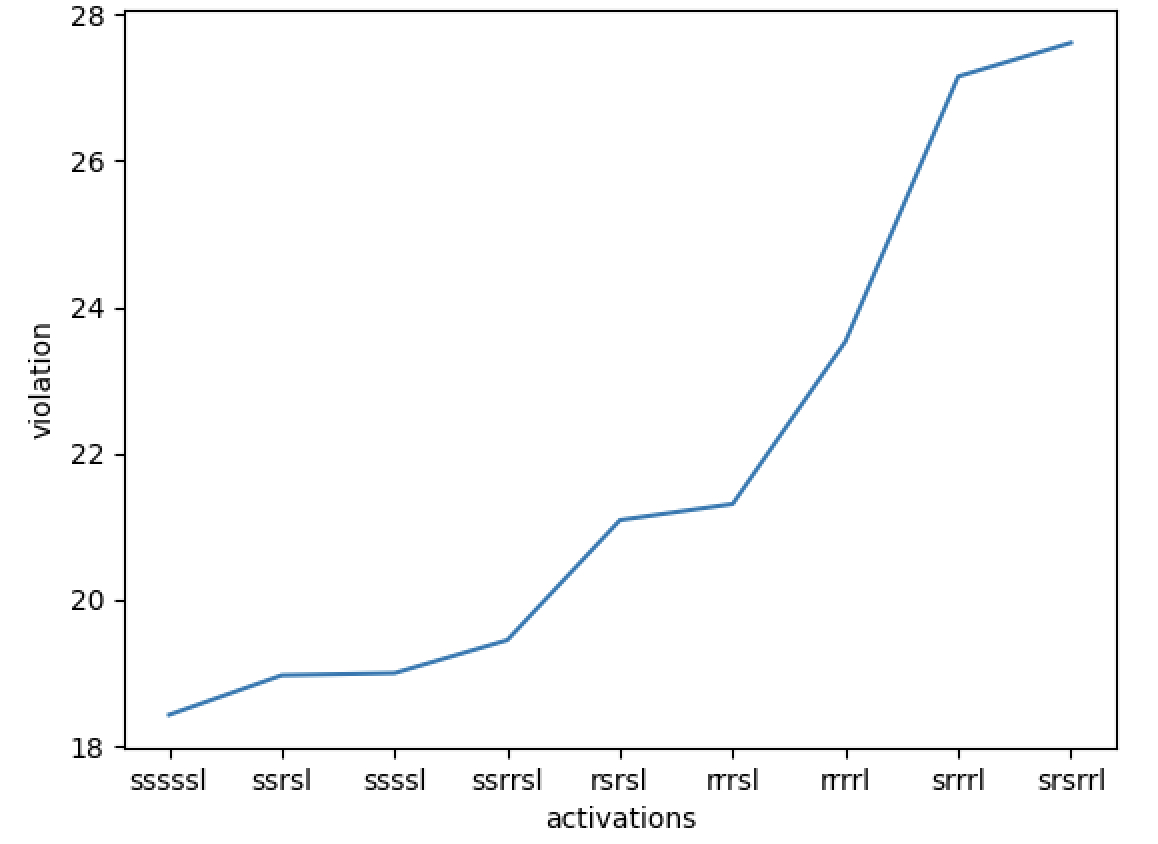}}
\caption{x-axis: activation structures of DeepReach;   y-axis: violation rate in percentage }
\label{violation}
\end{figure}

Whenever we apply Sine on the first and last layer of the DNN, violation rates are all below 20\%. With ReLU on the first layer and Sine on the last layer, violation rates are approximately 21\%. With ReLU on the first and last layer, violation rate is 23.5\%. With Sine on the first layer and ReLU on the last layer, violation rates are approximately 27\%.

The choice of activation on the first and last layer seems to have dominant effects on the performance, with subtle discrepancies for activation layers in the middle. We hypothesize that to learn substantial information about the heading($\theta$) parameter of the states, we need Sine on both the first and last layer to maintain good representation of the gradients. This discovery points to a new direction worthy of further experiments. 

\section{Limitations}

The key limitation of our approach is that, for various dynamical systems of different dimensions, we might discover disparate outcomes and patterns. Therefore, the empirical results shown in this paper cannot be well-generalized.

Secondly, only some combinations of Sine and ReLU are experimented with. We require a more systematic way of examining the influence of activation function on the accuracy of BRTs.


\section{Summary and Future Directions}

In this work, we explore how DeepReach performs on high-dimensional systems when using combinations of Sine and ReLU activation functions. We run experiments on 3D and 9D systems, and the results indicate that there exists a strong correlation between the number of Sinusoidal activation used and the performance of DeepReach on the 9D system. Moreover, the choice of activation on the first and last layer appears to have dominant effects on the performance of DeepReach than that of other layers. By using one more layer of Sine, we improved the violation rate from 19.0\% to 18.43\%.

In the future, we could explore other architectures with different hyperparameteres. Furthermore, to make learning-based reachability methods more applicable to complex systems in the real world, we may consider implementing a refined error correction model to optimize BRTs obtained from DeepReach with speed.








\section*{Acknowledgment}


The authors of this paper would like to express their appreciation to Professor Somil Bansal and Ph.D. candidate Hao Wang for their constructive feedback and suggestions that helped improve the quality of this paper.

\vspace{12pt}

\end{document}